# Capillary imbibition of monodisperse emulsions in confined microfluidic channels

Masoud Norouzi Darabad,[a] Sagnik Singha,[b] Jerzy Blawzdziewicz,[b,c] Siva A. Vanapalli [a] and Mark W. Vaughn [a]

We investigate imbibition of a monodisperse emulsion into a low-aspect ratio microfluidic channel with the height $h$ comparable to the droplet diameter $d$. For confinement ratio $d/h$ = 1.2, the tightly confined disk-like droplets in the channel move more slowly compared to the average suspension velocity. Behind the meniscus that drives the imbibition, there is a droplet-free region, separated from the suspension region by a sharp concentration front. The suspension exhibits strong droplet density and velocity fluctuations, but on average, the suspension domain remains uniform. For weaker confinement, $d/h$ = 0.65, the spherical droplets move faster than the average suspension flow, resulting in the formation of a dynamically unstable high-concentration region near the meniscus. We describe the macroscopic suspension dynamics using linear transport equations for the particle-phase flux and suspension flux that are driven by the local pressure gradient. A dipolar particle interaction model explains the observed large density and velocity fluctuations in terms of the dynamics of elongated particle clusters with different orientations.

## 1. Introduction

Capillary imbibition is a process in which a fluid driven by surface forces penetrates into a small pore. This process has various applications in science and engineering including lab-on-a-chip devices [1,2], oil recovery [3], and ink-jet printing [4]. In addition to its applications, capillary imbibition provides an unconventional method to investigate the microscale behaviour of dispersed-phase materials. By confining the dispersed-phase flow in a thin-rectangular capillary to achieve a two-dimensional-like geometry of a single droplet layer, the differences in the motion between phases is enhanced.

In order to describe the behaviour of an imbibed emulsion, we must first understand the imbibition of a simple liquid. The oldest report of this phenomena is found in the manuscripts of Leonardo da Vinci in 15th century[5], but the detailed dynamics were not understood until the 20th century. In 1921, EW Washburn[6] analysed the capillary imbibition of Newtonian fluids in a viscous dominated regime. He reported that the square of the penetration length is a linear function of imbibition time. In the early moments of imbibition, however, inertial forces may dominate, causing the dynamics to deviate from Washburn's predictions[7,8]. Furthermore, interpretation can be complicated by flow-induced changes in the contact angle[9]. The imbibition process is unsteady and temporal changes in the capillary number also affects, the dynamic contact angle and results in the deviations from Washburn's model[10,11].

For complex fluids, capillary imbibition has numerous biological and industrial applications. Examples include blood flow in microcapillaries [12-14], flow of polymers in nano-porous materials [15], fabrication of nano-composite materials [16] and optofluidics [17]. There are numerous studies for imbibition of single phase non-Newtonian fluids [14, 18-23], however there are only a few studies of capillary imbibition of suspensions. Zhou et. al. [12] studied the imbibition of blood suspensions and reported a failure of penetration due to capillary clogging as a result of segregation of red blood cells. Holloway et al. [24] studied the imbibition of a concentrated suspension of rigid spherical particles and reported that the dynamics of the imbibition obey Washburn's model, however the slope term is smaller than that predicted by using the effective viscosity of the suspension. They attributed this phenomenon to the accumulation of particles behind the meniscus and to the resulting non-uniformity of concentration in the axial direction. However, their analysis applies only to systems where the particle diameter $d$ is much smaller than the capillary diameter.

There does not appear to be a systematic investigation of the imbibition of suspensions under strong confinement conditions $d/h = O(1)$ (where $h$ is the smallest dimension of the channel). In contrast, the pressure driven flow of suspensions in narrow channels is well studied, especially in the Hele-Shaw geometry [25-29]. Pressure driven droplets/particles in Hele-Shaw cells exhibit fascinating collective behaviour such as vibrational modes in 1D ordered droplet arrays[30], dislocations [29, 31] fingering instabilities[31] in 2D ordered arrays, and Burger shock-waves developing in disordered droplet monolayers[26].

[a.] Dept. of Chemical Engineering, Texas Tech University, Box 43121 Lubbock, TX 79409, USA.
[b.] Dept. of Mechanical Engineering, Texas Tech University, Box 41021 Lubbock, TX 79409, USA.
[c.] Dept. of Physics and astronomy, Texas Tech University, Box 41501 Lubbock, TX 79409, USA.

The observed complex collective dynamics stems, primarily, from far-field dipolar Hele-Shaw hydrodynamic interactions between strongly confined particles[29, 30, 32]. The effects of higher-order contributions[33-35] on the microscale evolution[29, 35] and the macroscopic dynamics[34] have also been investigated. The presence of particles can either create or supress instabilities near the suspension front[36]. For example, the presence of particles in a suspension flow in a radial Hele-Shaw cell can result in viscous fingering.[36] Additionally, under strong confinements, particles can create a dense particle-band at the liquid-air interface, which can break-up due to the deformations of the interface.[37]

Here we study capillary imbibition of emulsions in thin, low aspect-ratio rectangular microfluidic channels. The droplet diameter $d$ is comparable to the channel height $h$, and the droplets form a monolayer in this Hele-Shaw geometry.

We consider two characteristic cases of the imbibition process, depending on the ratio between droplet diameter $d$ and the channel height $h$. In the first case, $d/h>1$, the droplets are flattened between the channel walls, and therefore have a circular disk-like shape. In the second case, $d/h<1$, the droplets remain spherical during the imbibition process. We explain how microscale droplet interactions result in distinct dynamics of the imbibition at macroscale in these two cases. For disk-like droplets, we observe formation of the droplet-free zone behind the meniscus with a sharp boundary between the droplet-free region and suspension region. For spherical droplets, as the particles accumulate behind the meniscus the densely packed accumulation region undergoes fingering instability that affects particle distribution and may cause strong deformation of the meniscus.

To shed light on the phenomenon observed in our capillary imbibition experiments, we use the linear transport equations to describe suspension flow in the channel, carry out microscopic analysis of particle cluster motion, and perform numerical simulations of particle motion using a dipolar-interaction model. We compare our experimental measurements with a benchmark model based on pressure driven-flow of hard-sphere monolayers in a parallel wall channel.

## 2. Materials and methods

### 2.1. Microfluidic device fabrication

Oil-in-water emulsions were prepared using a flow-focusing device [38, 39] with orifice width and length of 20 μm and 70 μm, respectively. The silicon wafer mould containing the designed flow-focusing devices was prepared using soft lithography techniques [40] and the resulting channels were 50 ± 0.4 μm in height. To make hydrophilic channels, the PDMS based chips were cut and punched with 500 μm biopsy punchers (Miltex) and bonded to glass slides (Fisher Scientific) by plasma bonding. The channels were then treated with water to maintain their hydrophilicity.

### 2.2. Emulsion preparation

Emulsions were prepared using water containing 2 wt% sodium dodecyl sulfate (SDS) (Sigma Aldrich) as the continuous phase and bromohexadecane (Sigma Aldrich) as the dispersed phase. The 2 wt% SDS corresponds to 8.6 times the critical micelle concentration[41]. The addition of the SDS is necessary to prevent droplet coalescence. The density of the dispersed phase, 0.999 g/mL at 25°C is close to water, so droplets are neutrally buoyant. Syringe pumps (PHD 2000, Harvard Apparatus) along with 1 ml plastic syringes are used to pump the fluids to the microfluidic device. To make monodisperse oil in water emulsions, bromohexadecane is pumped through the center channel of the flow focusing device[38] and then sheared in the orifice by SDS solution flowing from the side channels. The interfacial tension between bromohexadecane and SDS was measured to be 0.11 mN/m using the pendant drop method.

Stock emulsions were prepared with two different droplet diameters: d=60 μm and d=65 μm, both with $\phi$ = 0.50 oil volume fraction. The emulsions were then stored in microcentrifuge tubes (Fisher Scientific) and diluted to $\phi$ = 0.4, 0.3, 0.2, and 0.1 by adding 2 wt% SDS solution.

### 2.3. Experimental protocol

The system geometry is illustrated in Fig. 1a. Imbibition experiments were done using rectangular borosilicate glass capillaries (Vitrocom) that are $l$ = 5 cm long with a width to height ratio $w/h$ = 10. Two different sizes of capillaries are used, one with width $w$ = 500 μm and height $h$ = 50 μm and the other one with $w$ = 1000 μm and $h$ = 100 μm. Prior to conducting an experiment, the emulsions were carefully mixed by pipetting to remove non-homogeneity in droplets concentration due to slight creaming of the droplets. Then, using the pipette, an emulsion (approximately two times the volume of the capillary tube) is carefully placed in contact with the glass capillary inlet. The imbibition process starts immediately and is monitored using an inverted microscope (Nikon Eclipse- TiU) and a high-speed imaging camera (Phantom v710 12-bit, Vision Research). Videos were recorded at a maximum of 700 fps at 2X magnification with a pixel resolution of 10 μm per pixel. Experiments were recorded either at "Location 1" that covers the region from the capillary inlet to the position 1 cm away from the inlet (0 < x/l < 0.2) or "Location 2" that covers the region from 1.5 cm 2.5 cm away from the inlet (0.3 < x/l <0.5). These two locations are illustrated in the schematic of Fig. 1a.

### 2.4. Image processing

Image processing is implemented using a custom MATLAB (R2020a) program to determine the positions of the droplets and the meniscus. Prior to the imbibition process, a background image of the field of view is recorded. Since the droplet boundaries and the menisci appear as dark pixels, experimental frames and background image are inverted prior to background subtraction. Subtracted images are converted to a binary image with a threshold chosen to remove noise. The coordinates of the centre of the meniscus are used as its position. The results of the MATLAB program were verified by using ImageJ[42] to compare random samples to manual measurements.

To find the centroids of the droplets, the images are thresholded, and the connected regions identified using the `bwconncomp` function from the MATLAB (R2020a) image processing toolbox. A connected region with size comparable to a droplet and high circularity was identified as a potential droplet. The centroids of individual droplets can be used to extract their trajectories. We have created a custom MATLAB (R2020a) code that uses a Kalman filter[43] and the Hungarian algorithm[44] for multiple object tracking. The extracted tracks are inspected, and faulty tracks removed.

### 2.5. Area fraction measurements

It is essential to monitor the area fraction $\phi_a$ of the emulsion inside the capillary, since it can differ from the volume fraction $\phi$ of the original emulsion. These area fraction differences can stem from phenomena such as crowding of the droplets at the inlet of the capillary, the different flow rates of the droplets and the continuous phase fluid, and non-homogeneous droplet distribution in the original emulsion due to the creaming. The area fraction is measured directly by counting the number of droplets. See Fig. S1 and S2, ESI Section 1 for the evolution of the area fraction and the average area fraction for spherical and disk-like droplets. Using these data, we report an average area fraction $\bar{\phi}_a$ for each experiment.

### 2.6. Dimensionless numbers

The Reynolds number $Re$ and the capillary number $Ca$ for this system are defined as $Re = \rho u h/\mu$ and $Ca = \mu u/\sigma$ where $\rho$ is density, $\mu$ is viscosity, $\sigma$ is surface tension, and $u$ is the superficial velocity of the continuous phase. For a capillary with height $h$ = 50 μm, we have $Re \approx 1$ at location 1 and $Re \approx 0.3$ at location 2. The corresponding values for height $h$ = 100 μm are $Re \approx 2$ and $Re \approx 0.6$. These values were measured when the meniscus is in the middle of the field of view. The capillary number $Ca$ is less than 0.01 in all experiments.

According to the above values, inertial effects are relatively small, and Stokes-flow conditions approximately apply. The droplets are not deformed by the flow (except at the channel inlet for $d/h<2$) and are either circular or spherical in our experiments.

### 2.7. Simulations

To elucidate the observed droplet dynamics, we performed numerical simulations of a simplified system in which the droplets interact via the leading-order Hele–Shaw dipolar scatter flow fields[29, 30, 32], and the near-field effects are modelled using a short-range pairwise-additive repulsive potential, which prevents overlap. Previously, similar techniques were used for systems interacting via dipolar[26, 30], quadrupolar[27], and combined dipolar and quadrupolar[35] flows produced by the particles. To replicate the effect of side walls, we used a flow reflection method[45, 46]. For macroscopically homogeneous systems, periodic boundary conditions in the flow direction were applied. For inhomogeneous systems with a sharp concentration front, the leading portion of the suspension was simulated explicitly, and the trailing portion (far behind the front) was implicitly modelled by a repulsive potential acting in the flow direction and moving with the suspension velocity.

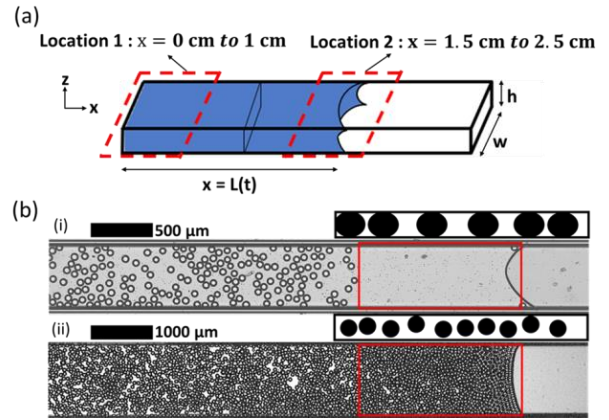

**Fig 1. System description.** (a) Schematic of the system. The glass capillaries were 5 cm long with an aspect ratio $h/w$ = 0.1. The imbibition experiments were recorded at two different locations, location 1: $x$ = 0 cm to 1 cm and location 2: $x$ = 1.5 cm to 2.5 cm. $L(t)$ is the penetration length of the meniscus at time t. (b) (i) Top-view image of capillary imbibition of disk-like droplets. Droplets with diameter $d$ = 60 μm imbibe into a capillary with $h$ = 50 μm and deform to a disk-like shape. The inset schematic shows the droplets in the $y$-$z$ plane. Disk-like droplets move slower than the continuous phase and a droplet-free region forms behind the interface. (ii) Top-view image of capillary Imbibition of spherical droplets. Droplets with $d$ = 65 μm imbibe into a capillary with $h$ = 100 μm. Spherical droplets flow faster than continuous phase and form a droplet-band behind the interface. The inset schematic shows the droplets in the y-z plane. The droplet-free region and droplet-band are shown by red rectangles.

## 3. Results and discussion

### 3.1. System description

A schematic of the experimental cell and two typical snapshots of the imbibition process are depicted in Fig. 1. In our analysis of the experiments, we use the coordinate system where the imbibition occurs in the axial direction $x$, and the channel width and height are along the $y$ and $z$ directions. The position of the meniscus at time $t$ is denoted by $L(t)$. The entrance location 1 and the mid-channel location 2 observation locations are marked by the dashed red lines (Fig. 1a).

We performed two sets of experiments: one set for strongly confined disk-like droplets and the other for moderately confined spherical droplets. In the experiments with strongly confined disk-like (Fig. 1b(i)), an emulsion of diameter $d$ = 60 μm droplets is imbibed into a capillary with height $h$ = 50 μm (confinement ratio $d/h$ = 1.2). After entering the capillary, the drops deform to a disk-like shape because they are squeezed between the top and

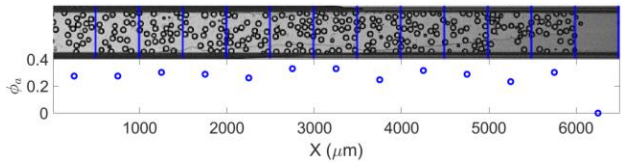

**Fig. 2. Distribution of disk-like droplets in the channel** A portion of the channel is divided into 13 equally sized sections and the area fraction in each section is measured. The first 12 sections (from 0 to 6000 μm) belong to the suspension region and the last section belongs to the droplet-free region. There is a sharp increase in the area fraction due to the sharp boundary between droplet-free and suspension region. The system exhibits considerable local drop density fluctuations, but no systematic area-fraction gradient was detected, consistent with qualitative observations of the particle distribution. The snapshot in this figure is taken from an experiment with $\bar{\phi}_a = 0.3$ at location 1.

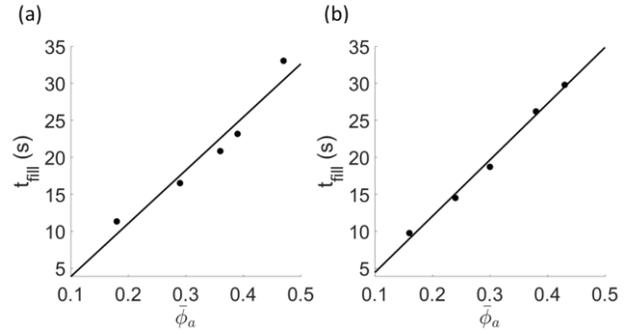

**Fig. 3. Filling time of capillary tube for disk-like droplets.** The imbibition time for an emulsion of disk-like droplets to fill the capillary tube measured at (a) Location 1 and (b) Location 2 of the channel. The filling time is an increasing function of the area fraction because presence of the droplets increases the viscous dissipation resulting in an increase in pressure drop that opposes the Laplace pressure of the bottom walls. For moderately confined spherical droplets (Fig. 1b(ii)), an emulsion of $d$ = 65 μm droplets are imbibed into a capillary with $h$ = 100 μm ($d/h$ = 0.65). The moderately confined droplets remain spherical, but they are large enough so that only a single layer forms in the capillary.

For both spherical and disk-like droplets, experimental measurements were conducted using five different volume fractions, ranging from 0 to 0.5. Two trials to monitor the imbibition process at locations 1 and 2 were performed for each volume fraction.

An image of a dispersion of strongly confined disk-like droplets imbibing into the capillary channel (Fig. 1b(i)) shows that two distinct regions develop during imbibition: there is a droplet-free region behind the meniscus, followed by a suspension region. This behaviour indicates that strongly confined droplets move slower than the continuous phase fluid. In contrast, Fig. 1b(ii) shows that moderately confined spherical droplets flow faster than the continuous phase. They catch up with the meniscus and create an adjacent dense droplet-band.

In the following two sections, we analyse the droplet distribution and the dynamic of the imbibition process in the disk-like and spherical droplet systems.

### 3.2. Imbibition of disk-like droplets

#### 3.2.1 Droplet distribution and overall imbibition dynamics

An examination of strongly confined systems depicted in Fig. 1b (i) and Fig. 2 reveals several characteristic features of the droplet distribution. First, the droplet-free and suspension regions are separated by a sharp concentration front. Second, the droplet distribution in the suspension undergoes considerable local fluctuations associated with formation of local droplet clusters. Moreover, large droplet-velocity fluctuations are observed. A similar behavior was seen for all area fractions investigated.

To determine the overall droplet distribution, the visible part of the channel is divided into 13 equal length sections of 500 μm, and the area fraction is measured in each section. Fig. 2 presents the result for an experiment with an average area fraction of $\bar{\phi}_a$=0.3. The first 12 sections (0 to 6000 μm) are from the suspension region, and the last section (6000 to 6500 μm) belongs to the droplet-free region. The bottom panel shows the measured value of the area fraction for each of the sections marked in the top panel. The results indicate considerable local droplet density fluctuations, but no systematic area-fraction gradient was detected, consistent with qualitative observations of the particle distribution.

To illustrate the overall imbibition dynamics, Fig. 3 shows the time $t_{\text{fill}}$ required for the emulsion to fill the entire capillary tube, plotted versus the average area fraction of the emulsion measured at locations 1 and 2. For experiments recorded at location 1 (Fig. 3a), $t_{\text{fill}}$ is determined by measuring the time interval between the meniscus formation at the inlet of the channel and the termination of droplet motion. For experiments recorded at location 2 (Fig. 3b), the imbibition time is evaluated from the observed time interval between the meniscus arrival at location 2 and termination of droplet motion. To this value, we add the estimated time it takes the meniscus to reach the imaged region; the estimate was obtained using Washburn analysis (details in ESI Section 2). We find that the results determined from observations at locations 1 and 2 are consistent.

As depicted in Fig. 3, $t_{\text{fill}}$ is an increasing function of the area fraction. The imbibition at higher area fractions is slower because the presence of droplets results in increased viscous dissipation and therefore, an increased pressure drop opposing the Laplace pressure of the meniscus. In the next section, we characterize the imbibition dynamics using appropriate transport coefficients.

#### 3.2.2 Constitutive transport coefficients of disk-like droplets

In the effective medium approximation, the suspension flow in the channel can be determined using a set of constitutive transport equations for the average particle velocity $\bar{U}$ and the volumetric suspension velocity $\bar{u}$:

$$\bar{U} = -\bar{v}^{tp}\nabla_{\parallel} p_s \qquad (3\text{-}1)$$

$$\bar{u} = -\bar{v}^{pp}\nabla_{||}\,p_s \qquad (3\text{-}2)$$

These two equations, written here for the axial components, link the local velocities $\bar{U}$ and $\bar{u}$ to the local pressure gradient $\nabla_{||}p_s$ via two linear transport coefficients $\bar{v}^{tp}$ and $\bar{v}^{pp}$. In addition, we introduce the transport equation

$$\bar{U} = \chi_c \bar{u} \qquad (3\text{-}3)$$

that describes particle convection from the average flow. According to these equations, the particle-phase mobility coefficient $\bar{v}^{tp}$, the effective permeability coefficient $\bar{v}^{pp}$ and the convective transport coefficient $\chi_c$ are related:

$$\bar{v}^{tp} = \chi_c\,\bar{v}^{pp} \qquad (3\text{-}4)$$

The transport coefficients $\bar{v}^{tp}$, $\bar{v}^{pp}$, and $\chi_c$ depend on the confinement ratio $d/h$, area fraction $\bar{\phi}_a$, and flow-induced local drop distribution. Our goal is to estimate the parameters $\bar{v}^{tp}$, $\bar{v}^{tp}$ and $\chi_c$ based on the capillary imbibition data.

Since the fluid is incompressible, the volumetric suspension velocity $\bar{u}$ is the same as the average fluid velocity in the droplet-free region $\bar{u}_0$

$$\bar{u} = \bar{u}_0 \qquad (3\text{-}5)$$

Thus, the suspension velocity $\bar{u}$ can be determined from the measurements of the meniscus velocity. The incompressibility condition eqn (3-5) also allows the pressure gradient in the suspension region, $\nabla_{||}p_s$, to be related to the pressure gradient in the clear fluid region, $\nabla_{||}\,p_c$,

$$\bar{v}^{pp}\nabla_{||}P_s = \bar{v}_0^{pp}\nabla_{||}p_c \qquad (3\text{-}6)$$

where $\bar{v}_0^{pp} = \frac{1}{12}\eta^{-1}H^2$ is the permeability of the channel filled with the clear continuous-phase fluid.

The total pressure drop in the channel is

$$\Delta P = -\big(l_c\nabla_{||}P_c + l_s\nabla_{||}\,P_s\big) \qquad (3\text{-}7)$$

where $l_c$ and $l_s$ are the instantaneous lengths of the droplet-free and droplet-dispersion regions and $\Delta P = 2\,\sigma\cos\theta/h$ is estimated from the Young-Laplace equation. The contact angle $\theta$ is not directly observable in our experiments, but we estimate it from single-phase imbibition experiments using 2 wt% SDS solution.

Using eqn. (3-6), and (3-7), we find the permeability coefficient of the suspension filled channel as

$$\frac{\bar{v}^{pp}}{\bar{v}_0^{pp}} = \frac{l_s}{l_{0c} - l_c} \qquad (3\text{-}8)$$

where $l_{0c} = (\bar{v}_0^{pp}\Delta P)/\bar{u}$ is the penetration length for which the clear fluid that moves with the velocity $\bar{u}$ in a single-phase experiment. The velocity $\bar{u}$ and the lengths of the clear fluid and suspension regions $l_c$ and $l_s$ are measured during the emulsion imbibition process (details in ESI Section 3) then used to evaluate the permeability coefficient $\bar{v}^{pp}$.

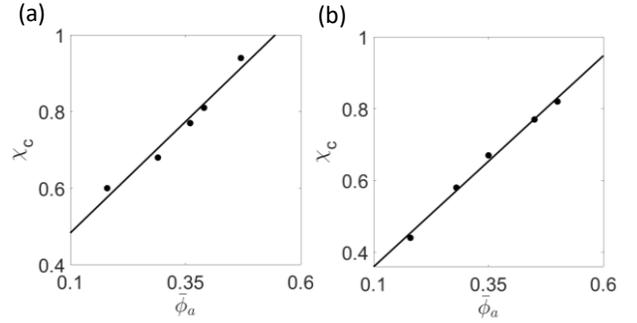

Fig 4. **Convective transport coefficient $\chi_c$ of disk-like droplets.** Convective transport coefficients χc of disk-like droplets vs area fraction for (a) Location 1 and (b) Location 2 of the channel. The coefficient χc increases with $\bar{\phi}_a$. This is because at higher area-fractions the particle phase produces stronger hydrodynamic resistance for the flow of the continuous-phase fluid in the space between the particles.

To determine the convective transport coefficient $\chi_c$ we use eqn (3-3). The average droplet velocity is evaluated from the motion of the suspension front—since there is no discernible macroscopic gradient of the area fraction $\bar{\phi}_a$ (Fig. 2), the front velocity is the same as the droplet-phase velocity $\bar{U}$. After evaluating the transport coefficients $\bar{v}^{pp}$ and $\chi_c$, the particle mobility $\bar{v}^{tp}$ is obtained using eqn (3-4). All transport coefficient values evaluated from subsequent video frames during a given experiment are averaged to report the final value.

Fig. 4 shows the convective particle transport coefficient $\chi_c$ vs the area fraction $\bar{\phi}_a$, evaluated from the measurements at location 1 (Fig. 4a) and location 2 (Fig. 4b). Location 1 (close to the entrance) tends to yield higher values of $\chi_c$ than location 2, even at the lowest area fractions considered. The observed location dependence of $\chi_c$ may stem from a decreased drop–wall lubrication friction associated with the higher droplet velocity at location 1. Other nonlinear effects not accounted for in our analysis (such as inertial effects and the evolution of particle distribution) may also affect $\chi_c$.

The results in Fig. 4 show that the convective transport coefficient $\chi_c$ increases with $\bar{\phi}_a$ and approaches values close to unity for higher area-fractions. This is because at higher area-fractions, the particle phase produces stronger hydrodynamic resistance for the flow of the continuous-phase fluid in the space between the particles. Therefore, both the fluid and the particles translate with the average velocity close to the overall suspension velocity.

This mechanism can be qualitatively modelled by approximating the particle phase as a porous medium through which the fluid phase flows with the permeability coefficient $\mu_s$. The particle medium moves with respect to the channel walls, with the friction coefficient $f$. By balancing the wall friction force with force resulting from fluid resistance, we obtain the expression (Details in the ESI Section 5)

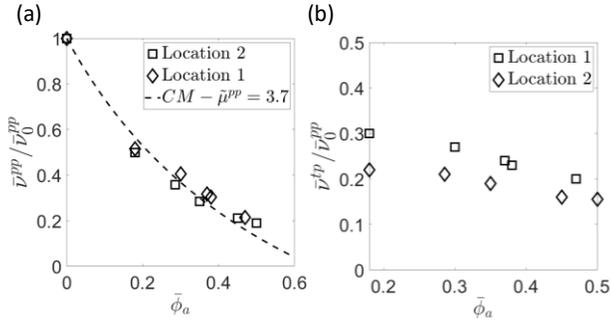
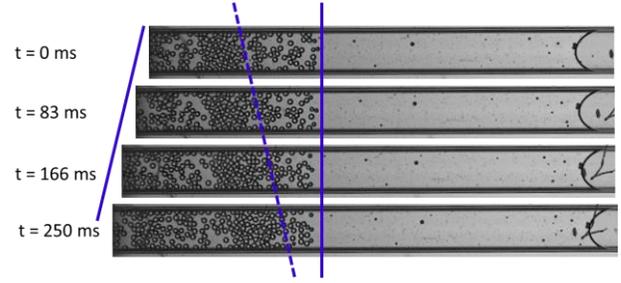

**Fig. 5. Transport coefficients of disk-like droplets.** (a) The effective permeability coeficinets $\bar{v}^{pp}$ of suspension of disk-like droplets vs $\bar{\phi}_a$. Square markers are for location 1 and diamond markers are for location 2. The dashed line is Clausius-Mossoti approximation where expansion coefficient $\tilde{\mu}_1 = 3.7$ is estimated from the experiment with the lowest area fraction. $\bar{v}^{pp}$ is a decaying function of $\bar{\phi}_a$. (b) The droplet mobility coefficient $\bar{v}^{tp}$ vs $\bar{\phi}_a$. The coefficient $\bar{v}^{tp}$ has a weaker dependence on the area fraction compared to $\bar{v}^{pp}$, because the convective transport coefficient $\chi_c$ is an increasing function of area fraction, partially compensating for the decay of $\bar{v}^{\wedge}pp$ in eqn (3-4).

**Fig. 6. Density wave propagation during the imbibition of disk-like droplets.** The wave velocity predicted by eqn (3-13) is greater than average droplet velocity. Therefore, the density waves associated with spontaneously occurring density fluctuations propagate faster than the average particle velocity. As a result, the distance between the front of the density wave (depicted by dashed line) and the front of the suspension region (solid vertical blue line) is decreasing.

.

$$\frac{1}{\chi_c} = 1 + (1 - \bar{\phi}_a)\bar{f} \qquad (3\text{-}9)$$

where $\bar{f} = f\mu_s/h$ is a dimensionless friction factor. An increase in the area fraction $\bar{\phi}_a$ leads to a decrease in the dimensionless friction factor $\bar{f}$ as well as its prefactor $1 - \bar{\phi}_a$, causing $\chi_c$ to grow towards unity.

Fig. 5a shows the values of $\bar{v}^{pp}$ versus the area fraction, evaluated at locations 1 and 2. As expected, the suspension permeability coefficient $\bar{v}^{pp}$ is a decaying function of the area fraction $\bar{\phi}_a$, because of the increased flow resistance at higher droplet concentrations. The dashed line in Fig. 5a is the Clausius-Mossoti approximation for permeability coefficients $\bar{v}^{pp}$ [47]

$$\bar{v}_{CM}^{pp} = \bar{v}_0^{pp}\frac{(1 - 0.5\,\tilde{\mu}_1^{pp}\phi_a)}{(1 + 0.5\,\tilde{\mu}_1^{pp}\phi_a)} \qquad (3\text{-}10)$$

where the low-density expansion coefficient $\tilde{\mu}_1^{pp} = 3.7$ is estimated from the experiment with the lowest area fraction (details in the ESI Section 4).

The droplet mobility coefficient $\bar{v}^{tp}$ evaluated from the measured transport coefficients $\chi_c$ and $\bar{v}^{pp}$ using eqn (3-4) is depicted in Fig. 5b. The coefficient $\bar{v}^{tp}$ has a weaker dependence on the area fraction compared to $\bar{v}^{pp}$, because the convective transport coefficient $\chi_c$ is an increasing function of area fraction, partially compensating for the decay of $\bar{v}^{pp}$ in eqn (3-4).

### 3.2.3 Formation of a sharp suspension front and the density fluctuation dynamics of disk-like droplets

During the capillary imbibition of disk-like droplets, a sharp front forms between the suspension domain and droplet-free region (Figs. 1b (i) and 2). We also observe that the density fluctuations in the suspension exhibit wave-like dynamics (Fig. 6 and ESI video).

Both the density-front formation and the propagating density waves can be explained using the convective transport eqn (3-3) and our measurements showing that the transport coefficient $\chi_c$ is an increasing function of $\bar{\phi}_a$, Fig. 4. Domains with higher area fraction move faster than lower-density regions. In a zone where the area fraction decreases, such as the leading end of the suspension region, the higher density portion of the suspension catches up with the lower density portion. This results in sharpening the area-fraction gradient and eventually forming a density front. For moderate amplitude density fluctuations, a similar mechanism produces density waves propagating in the forward direction, because the leading end of the dense region collects particles.

Quantitatively, front formation and wave propagation are described by the convective-transport equation for the suspension density. This equation, obtained by combining the constitutive relation eqn (3-3) with the continuity condition for the particle phase,

$$\frac{\partial \bar{\phi}_a}{\partial t} = -\nabla_{||}(\bar{\phi}_a\,\bar{U}) \qquad (3\text{-}11)$$

can be written in the form of the generalized inviscid Burgers equation,

$$\frac{\partial \bar{\phi}_a}{\partial t} = -c(\bar{\phi}_a)\nabla_{||}\bar{\phi}_a \qquad (3\text{-}12)$$

where

$$c(\bar{\phi}_a) = \left[\chi_c(\bar{\phi}_a) + \bar{\phi}_a\frac{d\chi_c(\bar{\phi}_a)}{d\bar{\phi}_a}\right]\bar{u} \qquad (3\text{-}13)$$

is the area-fraction-dependent velocity of the density waves. The Burgers equation eqn (3-12) predicts the formation of density shocks between the high- and low-density regions for systems where the wave velocity $c(\bar{\phi}_a)$ is an increasing function of the area fraction $\bar{\phi}_a$. This is the case for our experimental system. Shock formation can be demonstrated using the method of characteristics, and it results from the characteristics crossing. The observed sharp front between the suspension and clear-fluid regions is an example of a density shock predicted by the Burgers equation eqn (3-12).

We note that the wave velocity eqn (3-13) is greater than the convective velocity eqn (3-3). Therefore, the density waves associated with spontaneously occurring density fluctuations propagate faster than the average particle velocity and t the suspension front. This behavior is shown in Fig. 6, where the distance between the propagating low-density region and the front gradually decreases.

The observed wave-propagation direction relative to the average particle velocity in the quasi-2D imbibition system is opposite the one reported for 1-D drop trains[30]. For linear chains oriented parallel to the flow direction, the particle mobility decreases with the increasing particle density per unit length. The decrease of the mobility coefficient explains the backward direction of the propagating density wave.

For pressure driven flow of disk-like droplets in a microfluidic Hele–Shaw cell, density-wave propagation and shock formation was reported by Beatus, Tlusty, and Bar–Ziv.[26] Our analysis provides a generalization of their results in the context of the imbibition process.

### 3.2.4 Self-organized droplet clusters and their dynamics.

As shown in Figs. 7 and 8, droplets self-organize into a variety of transient local arrangements, which include short chains of different orientations, some elongated and some more compact, as well as high density regions separated by gaps of much lower particle density.

Our numerical simulations using the dipolar-interactions model described in section 2.7 capture the emergence of these transient structures, as depicted in Fig. 7. The simulations also replicate the sharp border between the suspension and clear fluid regions. Apart from the axisymmetric near-field repulsion, particles in our model interact only via the Hele–Shaw dipoles.

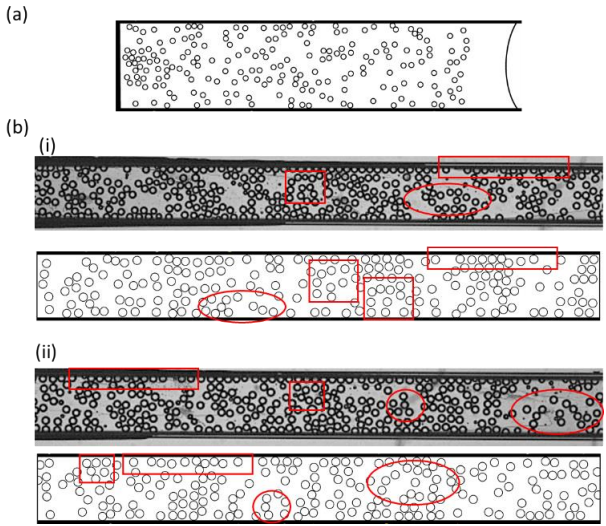

**Fig 7. Suspension model with dipolar Hele-Shaw hydrodynamic interactions.** (a) Similar microstructural features between simulation and experiment. The meniscus is drawn as a guide for the eye (b) Comparison between predicted microstructural features (bottom images) and experiments (top images). The same geometrical shape is used to highlight similar structural elements in experimental snapshots and simulations.

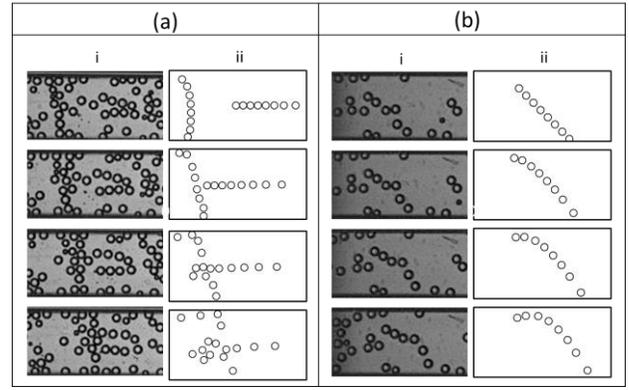

**Fig 8. Dynamics of longitudinal, transverse, and diagonally oriented droplet chains.** (a) The collision of a fast moving transverse droplet chain with a slow moving longitudinal chain resulting in a scattering of the droplets in (i) imbibition experiment, (ii) simulations based on dipolar interactions (b) The transverse motion of diagonal droplet chains toward the side walls in (i) capillary imbibition experiments, (ii) simulations based on dipolar interactions.

The agreement between the experimental and simulated microstructural features indicates that the dipolar hydrodynamic interactions provide the main organizing factor that leads to the emergence of the observed microstructural dynamics.

The chains generated during the imbibition process are usually short, with a typical length of several particles. They have relatively short lifetimes because of inherent instabilities[35] or collisions with other microstructural features. We note that isolated transverse chains are more stable[48] than longitudinal ones[35]. In general, the microstructure generated during the imbibition is much less ordered than the one seen for a system of share-driven deformable dropletss[49] for which dynamics are governed by the Hele-Shaw quadrupolar interactions instead of dipoles[27].

As illustrated in the time-lapse sequence of video images presented in Fig. 8a, chains and other elongated clusters with transverse orientation (across the channel) move faster than longitudinally oriented ones. Fig. 8b shows that diagonally oriented clusters have a nonzero transverse velocity component. Both the variation of the chain/cluster velocity with its orientation and the higher velocity of dense regions discussed in Sec. 3.2.1 contribute to the large particle velocity fluctuations that occur during the imbibition process, Fig. 9. The cluster velocity variation and its associated cluster collisions, Fig. 7a, also give rise to high density regions, enhancing the area-fraction fluctuations.

The Hele–Shaw dipolar model captures these features of droplet dynamics. The variation of the velocity and direction of cluster motion stems from the geometry of dipolar backflow patterns, as depicted in the schematic shown in Fig. 10, panels (ii). To further elucidate these phenomena, we approximate a particle chain as an elongated rigid body. In this simple approximation, chain motion is described by the mobility relation

$$\boldsymbol{V} = \boldsymbol{v}^c \cdot \boldsymbol{u} \tag{3-14}$$

where $\boldsymbol{u}$ is the velocity of the incident flow,

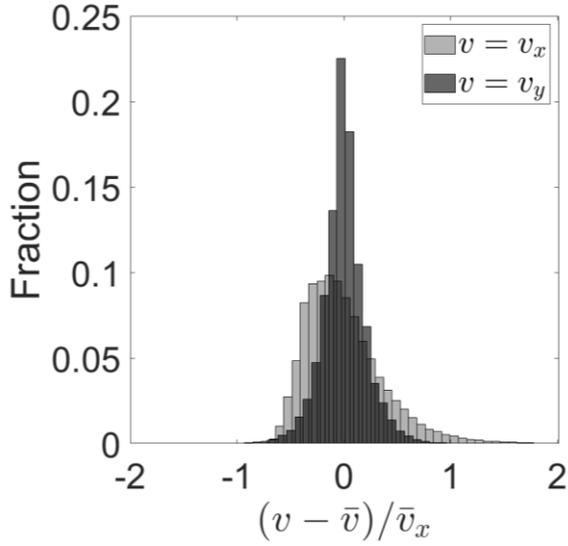

**Fig. 9 – Distribution of x and y component of the droplet velocity** The distribution of the velocity of the droplets in the x-direction $v_x$ (the flow direction) is wider than the distribution in the y-direction ($v_y$) (parallel to flow direction. This is due to the variation in the speed of droplet chain with different orientations as well as wave propagations discussed in Section 2.3.2.1.

$$\boldsymbol{v}^c = v_\parallel^c \hat{\boldsymbol{t}}\hat{\boldsymbol{t}} + v_\perp^c \hat{\boldsymbol{n}}\hat{\boldsymbol{n}} \tag{3-15}$$

is the chain mobility tensor, $\hat{\boldsymbol{t}}$ and $\hat{\boldsymbol{n}}$ are unit the vectors parallel and perpendicular to the chain, and $v_\parallel^c$ and $v_\perp^c$ are the longitudinal and transverse mobility components.

In strongly confined systems, the mobility tensor of a particle chain or an elongated particle cluster is strongly anisotropic[34, 50], with the transverse mobility significantly larger than the longitudinal mobility,

$$v_\perp^c > v_\parallel^c \tag{3-16}$$

In a quasi-2D system, the fluid cannot easily bypass a chain oriented normal to the flow direction, thus pushing the chain forward. In contrast, a longitudinal chain can be more easily bypassed by the flow, Fig. 10 panels (i).

Combining relations (3-14) and (3-15) we get the expressions
$$V_x = [v_\parallel^c \sin^2(\theta) + v_\perp^c \cos^2(\theta)]\bar{u} \tag{3-17a}$$
and
$$V_y = (v_\perp^c - v_\parallel^c)\sin(2\theta)\bar{u} \tag{3-17b}$$

for the longitudinal and transverse chain velocity components $V_x$ and $V_y$, where $\theta$ is the angle between the flow direction $\hat{\boldsymbol{e}}_x$ and the normal to the chain $\hat{\boldsymbol{n}}$. Taking into account the anisotropy of the chain mobility tensor, eqn (3-16), relation (3-17a) shows that chains with approximately transverse orientation $\theta \approx 0°$ move faster along the channel than the longitudinal ones ($\theta \approx 90°$). Moreover, diagonally oriented chains ($0°<\theta<90°$) have a nonzero transverse velocity component, predicted by eqn (3-17b). The experimentally observed chain behaviour is consistent with these theoretical results.

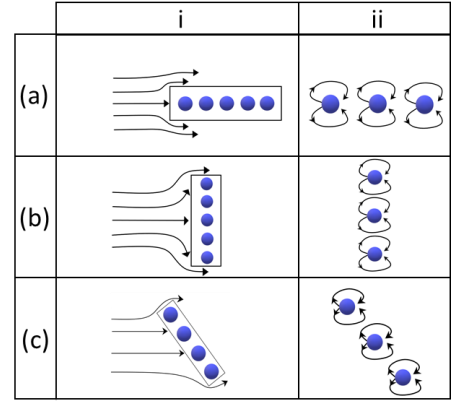

**Fig 10. Schematics showing the flow of fluid streamline and dipolar backflows in longitudinal, transverse and diagonal droplet chains.** (i) Fluid streamlines around (a) longitudinal, (b) transverse, and (c) diagonally oriented chains. (ii) Geometry of a dipolar backflow pattern for a) longitudinal (b) transverse (c) diagonally oriented chain.

### 3.3. Imbibition of spherical droplets

#### 3.3.1 Droplet distribution and overall imbibition dynamics

Imbibition of a monolayer of spherical droplets differs considerably from imbibition of strongly confined disk like droplets. The difference stems primarily from the opposite relation between the droplet-phase and suspension velocities $\bar{U}$ and $\bar{u}$. Spherical droplets move faster than the suspension (i.e., we have $\chi_c > 1$), whereas disk like droplets move slower ($\chi_c < 1$), as shown in Sec. 3.2. Moderately confined spherical droplets are faster than the average flow, because they reside in the midplane of the channel and therefore sample the streamlines with higher velocities. The focusing of droplets has been attributed to inertial migration [28] or deformation-based migration [51, 52].

As a result of the faster motion, spherical droplets reach the meniscus and form a dense droplet-band, as seen in the bottom panel of Fig. 1b and in Fig. 11. In contrast, there is a clear-fluid region behind the meniscus for the disk-like droplets. The dense region in the spherical drop system often has close-packed hexagonal domains and disordered regions. The border between these regions is unstable and diffuse. There is a systematic concentration droplet away from the meniscus, as quantitatively shown in the bottom panel of Fig. 11.

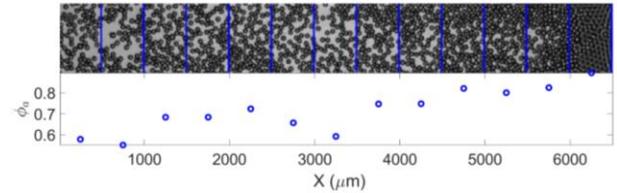

**Fig. 11 – Distribution of spherical droplets in the channel.** The area fraction of emulsion is measured in each of 13 equal divisions of a portion of the channel. The bottom panel shows the measured values of area fraction. We observe an axial gradient in the area fraction of the emulsion as well as large fluctuations in the area fraction. The snapshot is from an experiment with average area fraction for an emulsion with average area fraction of $\bar{\phi}_a$=0.53.

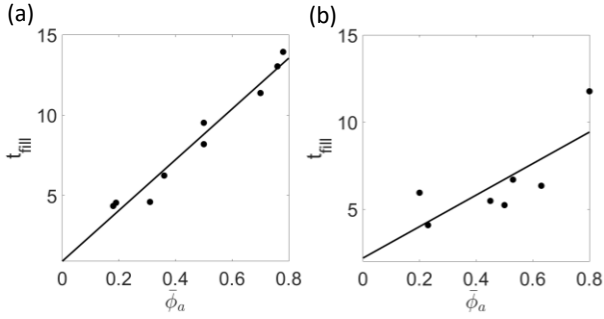

**Fig. 12. Filling time of capillary tube for spherical droplets.** The time required to fill the capillary tube with spherical emulsion, measured at (a) Location 1 and (b) Location 2 of the channel. The filling time increases with area fraction because the total friction force opposing the imbibition is larger at higher area fractions. The solid lines are least-square fits drawn as a guide for the eyes.

This gradually varying particle density distribution qualitatively differs from the one observed for disk-like drops, for which there is a sharp density front and no systematic density gradient. Moreover, according to the bottom panels of Figs. 2 and 11 the area fraction fluctuations are considerably larger for spherical droplets than those that occur for disk-like drops.

Fig. 12 shows the imbibition time for spherical droplets at locations 1 and 2 of the channel. The imbibition time is an increasing function of the area fraction, although this dependence is weaker than the one observed for disk-like droplets. There are also larger fluctuations from experiment to experiment. The fluctuations result from large emulsion density fluctuations and from the presence of the dense droplet region next to the meniscus, which affects the contact angle and the capillary pressure driving the imbibition. The density fluctuations and variation of the contact angle also cause inaccuracies in estimating the time for the meniscus to reach location 2; hence the difference of the imbibition time evaluated using observations at different locations.

### 3.3.2 Constitutive transport coefficients of spherical droplets

The dynamics of the suspension of spherical droplets is governed by the same set of constitutive equations, (3-1) to (3-3), as those that describe disk-like droplets. In the disk-like droplet imbibition, the presence of a uniform suspension region and a clear-fluid region separated by a sharp front allow determining the transport coefficients $\chi_c$, $\bar{v}^{pp}$, and $\bar{v}^{tp}$ with good accuracy. For spherical droplets, however, significant area-fraction nonuniformities and the lack of a front between the suspension and the clear fluid limits the ability to evaluate the transport coefficients. Therefore, the results are less complete than those given in Sec. 3.2.

We first consider the convective transport coefficient $\chi_c$. Its evaluation requires measurement of the volumetric suspension velocity $\bar{u}$ and the particle-phase velocity $\bar{U}$, eqn (3-3). As in the case of disk-like droplets, the suspension velocity $\bar{u}$ can be determined from the motion of the meniscus. However, the particle-phase velocity $\bar{U}$ is more difficult to measure because there is no suspension front moving with the velocity of the particle phase. In the low-area fraction experiments, $\bar{U}$ can be determined by particle tracking, but at higher area fractions, the droplets in high-density cluster are not easily tracked. Therefore, we provide results only for the lowest area fraction system, using data from experiments imaged at location 2.

Measurements yield $\chi_c = 1.26$ for $\bar{\phi}_a = 0.2$ and $\chi_c = 1.29$ for, $\bar{\phi}_a = 0.23$. As expected, $\chi_c > 1$, consistent with the qualitative observations. The experimentally determined value is lower by approximately 15% than the result $\chi_{0c} = 1.5$ for a point particle the midplane of the channel moving in the parabolic flow

$$u(z) = \frac{1}{6}\frac{z}{h}\left(\frac{z}{h} - 1\right)\bar{u} \tag{3-18}$$

where the channel walls are at $z = 0$ and $z = h$. The measured values are also lower, by less than 10%, than the value $\chi_c^{\text{Faxen}} = 1.39$ that is obtained by applying Faxen's law

$$\bar{U} = u + \frac{d^2}{24}\nabla^2 u \tag{3-19}$$

to an isolated solid sphere of diameter $d = 0.65h$ moving in the parabolic flow, eqn (3-18).

The drop velocity cannot exceed the point-particle velocity, and it is also unlikely to be larger than the Faxen's prediction, in which frictional forces due to the wall presence are neglected. We conclude that the coefficient $\chi_c(\bar{\phi}_a)$ remains nearly constant or decreases when the area fraction $\bar{\phi}_a$ is increased. This conjecture is consistent with the theoretical calculations for a hard-sphere monolayer, which has very weak area fraction dependence of $\chi_c$ (Sec. 4.2). It follows that the suspension of spherical drops does not show shocks and density-wave propagation that are observed for disk-like droplets.

Evaluation of the transport coefficient $\bar{v}^{pp}$ requires measuring the volumetric suspension velocity $\bar{u}$ and determining the local pressure gradient in a uniform suspension region. While $\bar{u}$ can be obtained from the meniscus velocity, the local pressure gradient cannot be evaluated from our experimental data because of suspension nonuniformities. Therefore, we calculate the average value of $\bar{v}^{pp}$ for the entire channel:

$$\bar{u} = \bar{v}^{pp}_{\text{avg}}\nabla_\parallel P_{\text{avg}} \tag{3-20}$$

where $\nabla_\parallel P_{\text{avg}} = \Delta P/L$ is the average pressure gradient, with the pressure drop $\Delta P$ evaluated from the Young-Laplace equation. This is similar to treatment of disk-like droplets.

The average permeability coefficient $\bar{v}^{pp}_{\text{avg}}$, based on measurements performed at locations 1 and 2, is plotted in Fig. 13 versus the average area fraction. The data show that $\bar{v}^{pp}_{\text{avg}}$ is a decaying function of the area fraction. As discussed in Sec. 4.2, the decay is much stronger than the one expected according to a comparison with results for a hard-sphere monolayer. Based on an analysis of high magnification images (Fig. 14), we attribute this discrepancy to displacements of highly packed drops out of the midplane of the channel. The displaced drops can create significant friction with the walls, hindering suspension flow. Our average results thus imply a high permeability contrast between low-density and nearly close-packed domains of the channel. By similar arguments, we also expect that the coefficient $\chi_c$ may significantly decrease at high area fractions.

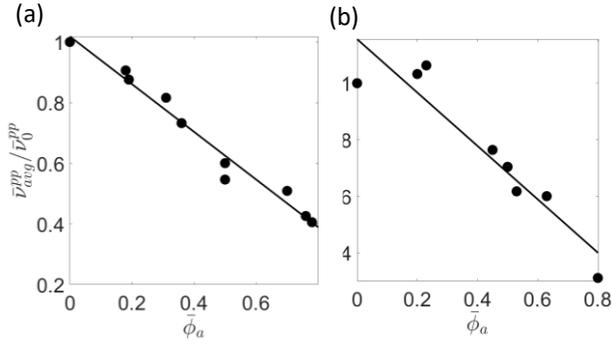

**Fig. 13. Average suspension permeability coeficient $\bar{v}^{pp}_{avg}$ for spherical droplets.** The measured values of $\bar{v}^{pp}_{avg}$ for disk-like droplets at (a) location 1, and (b) location 2. $\bar{v}^{pp}_{avg}$ is a decaying funcion of area fraction. However, the decay is much stronger compared to numerical sumulations for hard-sphere monolayers (Section 4.2).

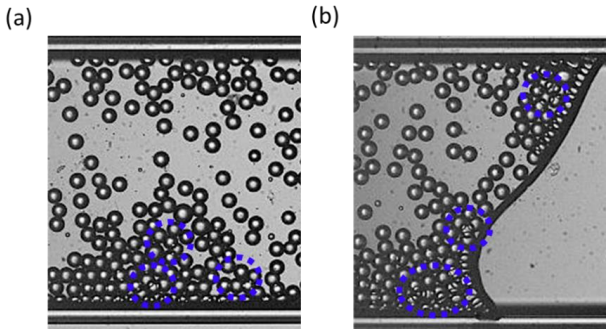

**Fig. 14. High magnification images of dense regions formed during the imbibition of spherical droplets.** High magnification images (4X) of dense regions formed during the imbibition of spherical droplets at (a) upstream of the meniscus (b) at the meniscus. The regions marked by blue circles show droplets with overlapping boundaries, indicating the displacement of the droplets out of midplane.

### 3.3.3 Suspension instability and interface deformations during the imbibition of spherical droplets

Strong fluctuations of the area fraction of spherical droplets and the existence of a systematic area-fraction gradient in the axial direction, stem from the instability of the dense droplet-band, which is locally displaced by less-dense suspension. An example of this instability is presented in the time-lapse image sequence depicted in Fig. 15.

The observed instability is analogous to the Saffman–Taylor viscous fingering instability[53] in a Hele–Shaw cell. The fingering occurs when a high viscosity fluid is displaced by a fluid with lower viscosity. Since the parts of the channel filled with the low-viscosity fluid are more permeable, fluid flows more easily in these domains, creating low-viscosity fingers. A similar behavior occurs here, where low droplet area-fraction regions have a larger channel permeability coefficient $\bar{v}^{pp}$ than the high area-fraction regions. See the average results shown in Fig. 13.

Hydrodynamic stresses associated with the motion of low-density fingers can result in meniscus deformation, Fig. 15. The deformation in turn facilitates finger formation because it breaks the hexagonally ordered suspension structure in the closely packed region near the suspension front. Closely packed ordered emulsion has a nonzero yield stress, so development of an ordered structure may prevent finger formation, allowing a buildup of a large dense domain. This behavior is seen in Fig. S7 in the ESI, which shows results at a higher confinement ratio. Formation of hexagonal structures is also predicted by our numerical Hele–Shaw dipolar model with meniscus approximated using a repulsive force acting in the direction opposite to the flow, Fig. 16.

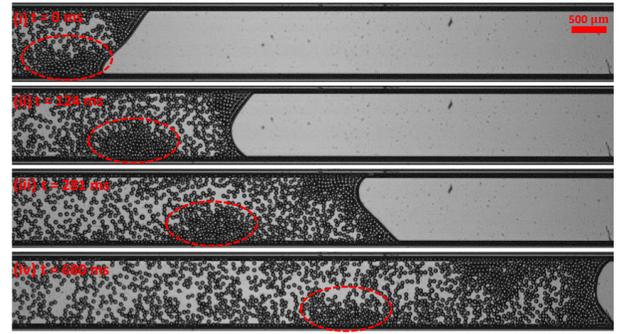

**Fig. 15. Fingering instability and interface deformations during the imbibition of spherical droplets.** Time-lapsed images from an experiment with $\bar{\phi}_a = 0.45$ at location 2, showing the fingering instability and deformation of the meniscus. The fingering instability and deformations in the interface results in the separation of dense droplet cluster from the close-packed droplet band migrating away from the meniscus and gradually disappears due to random droplet motions.

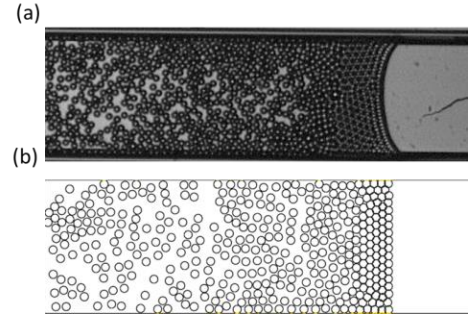

**Fig. 16. Hele-Shaw dipolar model predicts the formation of hexagonally packed droplet-band.** Qualitative comparison between the droplet band formed in an experimnt with $\bar{\phi}_a = 0.53$ at location 2 and a droplet-band predicted by Hele-Shaw dipolar model. The meniscus is simulated by using a repulsive force that acts in oppisite to flow direction.

When a low-density finger approaches the meniscus it usually turns towards one of the side walls of the channel. After touching the wall, the finger causes separation of a dense droplet cluster from the close-packed droplet band near the meniscus. As seen from the time-lapse images presented in Fig. 11, the separated clusters migrate away from the meniscus, gradually decaying from random droplet motions. This process periodically repeats, causing the observed large area- fraction fluctuations and formation of a systematic area-fraction gradient seen in Fig. 11.

Fingering instabilities were reported in pressure driven flow of a suspension in a radial Hele-Shaw cell[36, 37]. Our results demonstrate that the instability also occurs during imbibition and that it can significantly affect the imbibition dynamics.

## 4. Discussion and conclusions

### 4.1 Qualitative comparison between dynamics of dense clusters of disk-like and spherical droplet systems.

The results presented in Sec. 3.2 and 3.3 reveal significant differences between the imbibition hydrodynamics in the disk-like- and spherical-droplet systems. The observed differences are most evident near the meniscus, where either clear-fluid or densely-packed region forms. However, we also found fundamental differences in the mesoscale droplet-distribution dynamics away from the meniscus.

For disk-like droplets, there is a large variation between the velocities of longitudinally and transversely oriented chains. Such variation is less pronounced for spherical droplets, consistent with the fact that the mobility tensor (eqn (3-15)) is less anisotropic for weaker confinement [33, 34]. We also note that transverse chains of spherical droplets are less stable than the transverse chains/clusters of strongly confined disk-like droplets, as predicted by the analysis of chain stability at different confinement [48].

Dense clusters of disk-like droplets often span the width of the channel and exhibit density wave dynamics, as discussed in Sec. 3.2.3. Both the droplets in these clusters and the waves move faster than the drops in the low-density regions. In contrast, for spherical droplets, dense clusters usually span only part of the channel width and move slower than the low-density regions. Moreover, no density waves are observed in the spherical-droplet system.

The origin of distinct dynamics of density fluctuations is explained in the schematic shown in Fig. 17. Due to the flow continuity condition, the dynamics of transversely oriented clusters of disk-like-droplets (Fig. 17a) are primarily controlled by the convective trasport equation (3-3). The drop density varies mainly in the longitudinal direction. Because of the continuity condition, the average suspension velocity is the same in the region behind a cluster, in the cluster, and in front of the cluster. The convective transport coefficient has the largest value in the dense region, $\chi_{c1}, \chi_{c3} < \chi_{c2}$, so one concludes that the dense cluster moves faster than the lower area-fraction regions ( $\bar{U}_1, \bar{U}_2 < \bar{U}_3$).

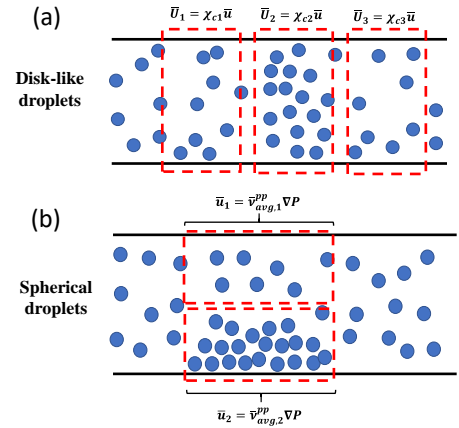

Fig. 17. Comparing the dynamics of dense droplet clusters for disk-like and spherical droplet systems. (a) The dense clusters of disk-like droplets usually span the whole width of the channel and its dynamics is governed by $\chi_c$ resulting in faster motion of the dense region (b) The dense cluster of spherical droplets do not span the whole width of channel and their dynamics is governed by $\bar{v}^{pp}_{avg}$ resulting in a slower motion of the dense region.

In a system of spherical droplets, the clusters that separate from the dense droplet-band do not span the entire width of the channel. As explained in Fig. 17b, their motion relative to the neighboring low-density regions is thus governed by the mobility relations (3-1) and (3.2). Since the pressure gradient in the cluster and a latelarly adjacent low-density domain are approxiamtely the same, and the dense region has a lower permeability coefficient $\bar{v}^{pp}_{avg,1} < \bar{v}^{pp}_{avg,2}$, it moves slower than the low-density region ( $\bar{u}_2 < \bar{u}_1$). Noting that the convective transport coefficient $\chi_c$ is reduced in dense clusters due to out-of-plane particle displacements, the particle velocity is further diminished. As a consequence, a significant motion of clusters away from meniscus is generated. This backward motion results in particle transport away from the meniscus, balancing the forward motion of droplets in low density regions, and preventing a continuous buildup of the densely packed suspension domain behind the meniscus.

### 4.2 Comparing the experimetal measurements with numerical simulations for confined hard-spheres.

In this section we compare the experimentally determined transport coefficients $\chi_c$, $\bar{v}^{pp}$, and $\bar{v}^{tp}$ to our benchmark calculations for a confined hard-sphere monolayer driven by a pressure gradient in a parallel-wall channel. The benchmark calculations were performed using the Cartesian representation method[33] for a hexagonal lattice of spherical particles moving in the midplane of the channel. The details of our technique along with a theoretical analysis of particle transport in parallel-wall channels are provided in[47]. The calculations are performed for a hexagonal lattice, while the experimental systems are mostly disordered. However, based on the comparison between hexagonal-lattice and square-lattice results, we expect that the microscale particle distribution does not strongly affect transport coefficients, as long as there are no large area fraction fluctuations (which in our calculations are not present).

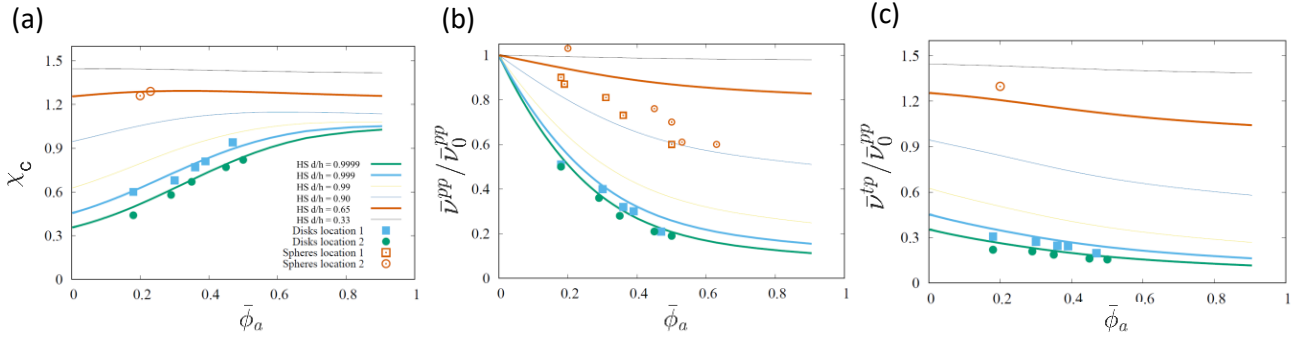

**Fig. 18. Experimental transport coefficients $\chi_c$, $\bar{v}_{pp}$, and $\bar{v}^{tp}$ compared with numerical simulations for confined hard spheres.** Lines represent the values from numerical simulations based-on confined hard-spheres and symbols represent the experimental measurements for (a) $\chi_c$, (b) $\bar{v}^{pp}$, and (c) $\bar{v}^{tp}$. For disk-like droplets, the experimental measurements match well with the hard-sphere results for $h/d = 0.999$ and $h/d = 0.9999$. For spherical-droplets measurements agree only at lower area fractions.

The theoretical hard-sphere results for $\chi_c$, $\bar{v}^{pp}$, and $\bar{v}^{tp}$ at confinement rations ranging from $d/h = 0.33$ to $d/h = 0.9999$ are depicted in Fig. 18 (lines) along with the corresponding experimental data for disk-like and spherical droplet systems (symbols). The results are plotted vs the area fraction $\bar{\phi}_a$, and each line represents calculation for a different confinement ratio.

We find that the measured transport coefficients for disk-like droplets quantitatively agree with the theoretical hard-sphere calculations for a strongly confined particle monolayer. Specifically, the measurements performed at locations 1 and 2 can be mapped onto the hard-sphere results for $h/d = 0.999$ and $h/d = 0.9999$, respectively. The spherical-droplet system overall agreement is less satisfactory, except at the lowest area fraction $\bar{\phi}_a \approx 0.2$, where the measured convective-transport coefficient $\chi_c$ agrees well with the hard-sphere results for the same confined ratio as the one used in our experiments. The overall agreement between experiment and benchmark calculations indicates particle motion in both the hard-sphere and droplet systems is governed by similar hydrodynamic mechanisms.

As argued in Sec. 3.2.2 and the ESI, section 5, particle dynamics at strong confinements is controlled by the competition between the pressure force pushing the particles along the channel and the particle–wall hydrodynamic friction. Moreover, the fluid motion relative to the particle phase is governed by the inter-phase permeability coefficient $\mu_s$, which depends on particle spacing.

For disk-like droplets, the particle-wall friction is generated in the thin fluid film between the wall and the drop interface, and for highly confined rigid spheres, it is produced in the lubrication layer in the near-contact region. Since the hard-sphere lubrication region is much smaller than the flattened-droplet film, we need to use very strong confinements to match the theoretical results to the experimental ones.

The confinement ratio $h/d$ required to match the results obtained from measurements at location 1 is smaller than the confinement ratio needed for location 2. This is consistent with our conjecture that the location-dependence of the measured values of the transport coefficients stems from the increased film thickness at the beginning of the imbibition process due to the faster drop motion. Similar area-fraction dependence of the droplet and hard-sphere transport coefficients seen in Fig. 18 suggests that the backflow associated with dipolar Hele-Shaw interactions is a key factor in suspension dynamics for both systems.

For spherical droplets in the moderately confined system, we find a good agreement between the experimental and theoretical results only at low area fractions. The agreement is especially good for the convective transport coefficient $\chi_c$, for which accurate particle-tracking results are available for the area fraction $\bar{\phi}_a \approx 0.2$.

At higher area fractions, experimental results are available only for the effective transport coefficient $\bar{v}^{pp}_{\text{avg}}$, estimated by averaging over the entire non-uniform suspension region. We find that the theoretical and experimental results significantly differ at higher values of fraction $\bar{\phi}_a$. According to our theoretical results, the coefficient $\bar{v}^{pp}$ decreases by less than 10% in area-fraction domain for which experimental results are available, whereas the experimentally determined values of $\bar{v}^{pp}_{\text{avg}}$, decrease by almost 40%. This significant difference between the experimental and theoretical results stems, most likely, from the displacements of spherical drops out of the midplane of the channel in high area-fraction regions, as discussed in Sec. 3.3.2.

### 4.3 Conclusions.

The presented results demonstrate rich phenomenology of the emulsion imbibition process under moderate and strong confinements. In particular we have shown that the drop distribution can be controlled by the degree of confinement. Under strong confinement where the droplets deform to a disk-like shape, we observe formation of droplet-free region behind the meniscus. Moreover, the suspension region has a sharp front boundary due to the presence of shockwaves. Droplets in the suspension region show strong velocity and arae fraction fluctuations with density wave propagations.

During the imbibition of moderately confined spherical droplets, a dense droplet-band forms behind the meniscus that undergoes fingering instability. As opposed to disk-like droplets there is no sharp boundary between droplet-band and suspension region. However, our experiments with spherical droplets at stronger confinements shows formation of closely-packed dense regions that can prevent fingering instability.

## Acknowledgements

Acknowledgement is made to the donors of the American Chemical Society Petroleum Research Fund for support of M.W.V and M.N in this research. S.S. and J.B. were supported by NSF Grant CBET 1603627. We thank Purushottam Soni for the surface tension measurement.

**Supplementary materials**

**1 - Area fraction measurements during the imbibition of disk-like and spherical droplets:**

Figs. S1 and S2 show how the area fraction of emulsion in field-of-view changes during the imbibition process. This is measured by directly counting the number of the droplets during the imbibition process.

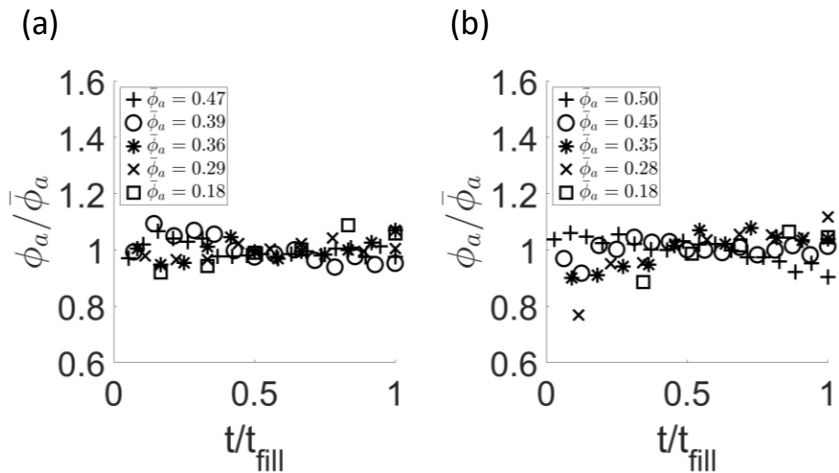

**Figure S1. Area fraction of disk-like droplets.** The observed area fraction of disk-like droplets during capillary imbibition for (a) Location 1 and (b) Location 2. The area fraction of each experiment is normalized with the average value observed in that particular experiment. The area fractions stay close to the average values throughout the imbibition.

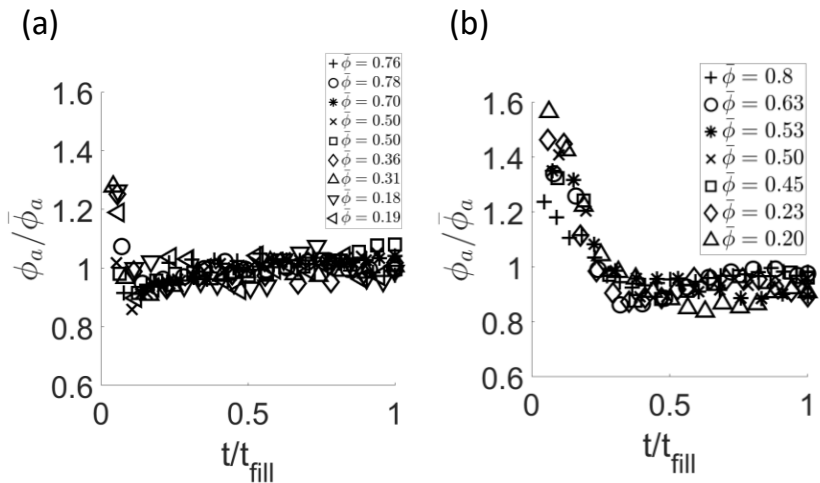

**Figure S2. Area fraction of spherical droplets.** The observed area fraction of disk-like droplets during capillary imbibition for (a) Location 1 and (b) Location 2. The area fraction of each experiment is normalized with the average value observed in that particular experiment. The area fractions stay close to the average values throughout the imbibition.

## 2 - Applicability of Washburn analysis to capillary imbibition of emulsions

According to Washburn's analysis[1], during the capillary imbibition of Newtonian fluids, the square of penetration length $L(t)^2$ is a linear function of time with a slope that depends on the channel and liquid properties. Washburn law is not valid very close to the inlet when the viscous boundary layer is not fully developed, and inertial forces dominate the imbibition process. In the inertial region, the position of meniscus $L(t)$ is a linear function of time [2]. So, the small portion of the experimental data from location one (from the inlet of the channel up to 5 times the width of the channel) is neglected to make sure that the imbibition process is in a viscous-dominated regime.

Suspensions are not generally considered Newtonian fluids, but this linear dependence between $L(t)^2$ and t is valid, even for concentrated suspension [3]. Motivated by this fact, the Washburn equation's applicability for the case of spherical and disk-like droplets is studied here.

Washburn [4] originally studied the imbibition process in circular cross-section channels, but his model can be modified for our system. Since the width of capillaries used in this experiment is 10 times larger than the height, it is reasonable to assume that the velocity gradient in the width direction is negligible with respect to velocity gradient in the height direction and approximation geometry a parallel wall channel.

By this assumption, the Naiver-Stokes equation simplifies to:

$$\frac{\partial P}{\partial x} = -\eta \frac{\partial^2 v}{\partial z^2} \tag{S1}$$

the pressure gradient at time $t$ when the liquid has traveled length $L(t)$ is:

$$\frac{\partial P}{\partial x} = \frac{\Delta p}{L(t)} = \frac{2\sigma}{L(t)} \cos\theta \left(\frac{1}{w} + \frac{1}{h}\right) \sim \frac{2\sigma}{L(t)h} \tag{S2}$$

Here $\Delta P$ is calculated by the Young-Laplace equation in which $\sigma$ is the surface tension, and $\theta$ is the contact angle that is assumed to be constant and equal to zero during the imbibition process. Also, the contribution of meniscus curvature in the *y*-direction is neglected compared to curvature in *the z*-direction. Substituting Eq. S2 in Eq. S1 and integrating will result in:

$$L^2(t) = \frac{h\sigma}{3\eta}t = St \qquad \text{(S3)}$$

Eq. S3 is the modified version of the Washburn equation for the case of parallel wall geometry. Experiments with single-phase liquids confirm the validity of this approximation for the capillary tubes used in this study.

Figure S3 a and b shows the square of penetration length of meniscus normalized by the width of the channel $\left(\frac{L(t)}{W}\right)^2$ as a function of time for different average area fraction of the disk-like emulsion in location 1 and 2. Markers show the experimental measurements and solid lines are the least-squares fit to the data. The dashed lines show the experiment with single phase SDS solutions. The data shows a linear trend for $L(t)^2$ versus time with $R^2 > 0.99$ for all the experiment. and confirms the validity of Washburn analysis for both locations in the capillary tube, even for the most concentrated emulsion. Also, it confirms that the contact angle is not changing severely during the imbibition. Otherwise, it would appear as a change in the slope during imbibition.

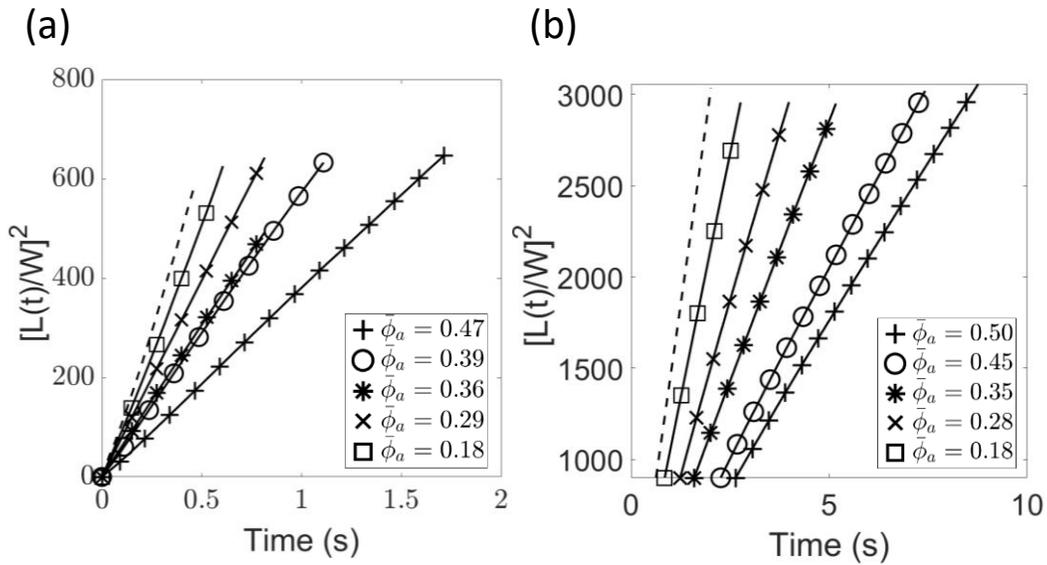

**Figure S3. Applicability of Wasbhurn to disk-like droplets** The $\left(\frac{L(t)}{W}\right)^2$ versus time for disk-like droplets at (a) location 1 and (b) location 2. The markers show the experimental measurements, the solid-lines are least-squares fit to the data and

For experiments conducted at location 2, the exact start time of the imbibition is not known. However, it can be estimated by finding the amount of shift it is needed to make the Washburn's fits pass through the origin. Using this estimation, we can calculate the time meniscus has traveled until it has reached location 2.

The same analysis has been conducted for spherical droplets. Results are summarized in Fig. S4:

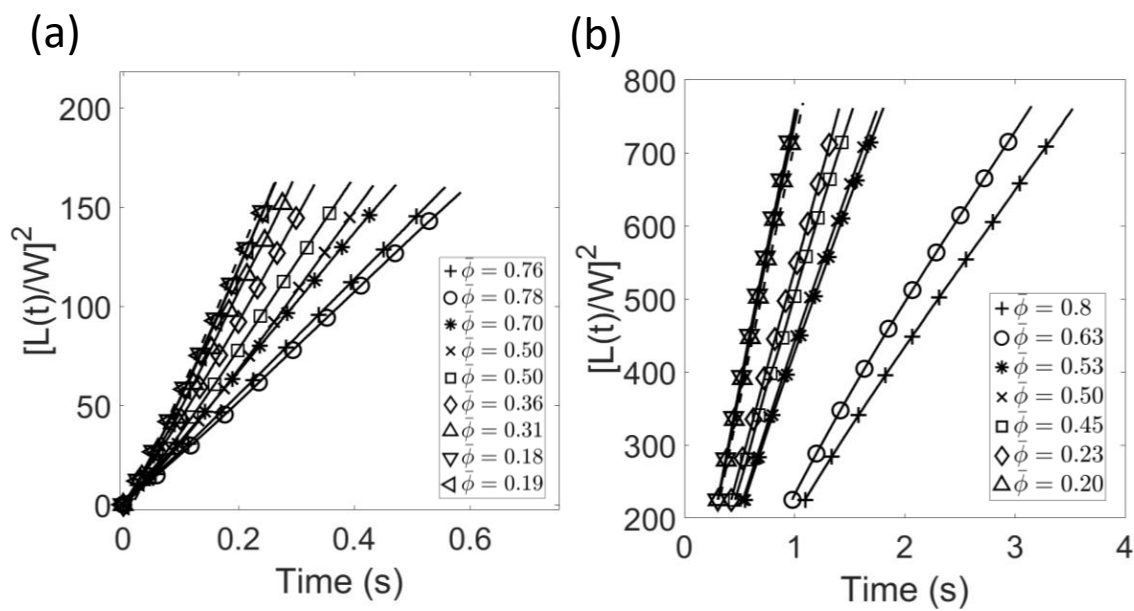

**Figure S4. Applicability of Wasbhurn to disk-like droplets** The $\left(\frac{L(t)}{W}\right)^2$ versus time for spherical droplets at (a) location 1 and (b) location 2. The markers show the experimental measurements, the solid-lines are least-squares fit to the data and the dashed-line is the least-squares fit to data from single-phase SDS experimnets.

## 3 - Measured values of $L(t)$, $l_c(t)$, and $l_s(t)$ for disk-like droplets used in the calculation of $\bar{v}^{pp}$, $\bar{v}^{tp}$ and $\chi_c$:

In order to calculate the values of $v^{pp}$ and $v^{tp}$ and $\chi_c$ from capillary imbibition data, we need instantaneous values of $L(t), l_c(t)$, and $l_s(t)$. The exact definition of these are depicted in the schematic in supplementary Fig. S5.

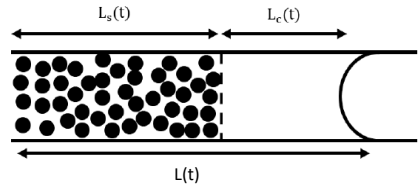

**Figure S5. Schematic for capillary imbibition of disk-like droplets.** L(t) is the imbibition length at time t, $L_s(t)$ is the length of suspension region at time t and $L_c(t)$ is the length of clear fluid region.

Fig. S6. Shows the measured values of $L(t), l_c(t)$, and $l_s(t)$ for capillary imbibition of disk-like droplets:

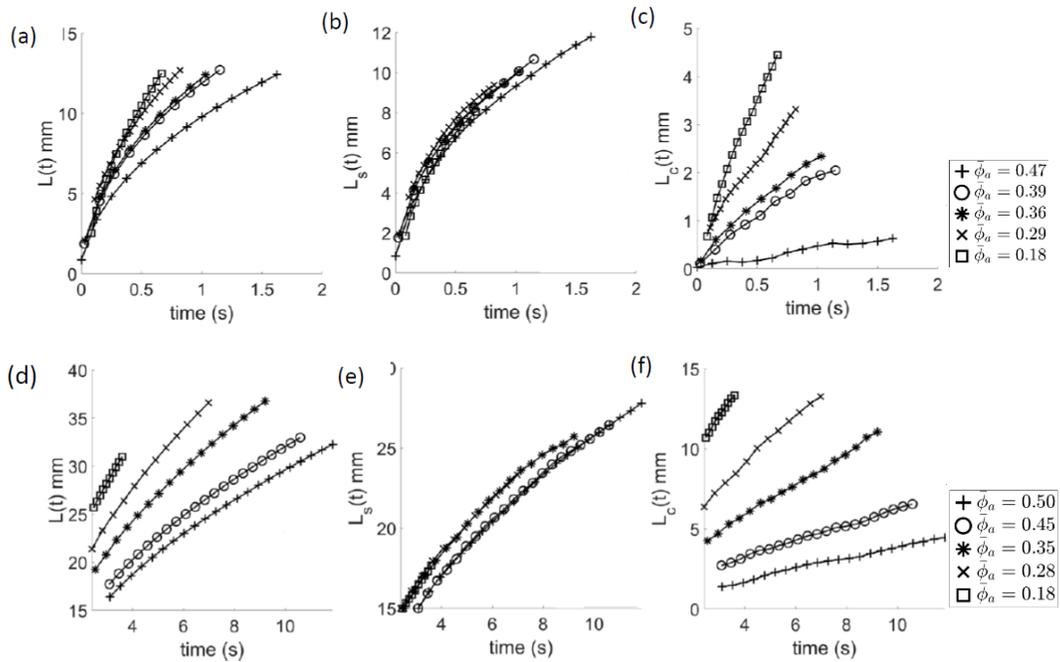

**Figure S6:** Evolution of $L(t)$, $Ls(t)$ and $Lc(t)$ versus time for different volume fraction. Top panel is for location 1 and bottom panel is for location 2

## 4 - Clausius-Mossotti approximation

The expression for Clausius-Mossitti approximation is given as[5]:

$$\frac{\bar{v}^{pp}}{\bar{v}_0^{pp}} = \frac{1 - \frac{1}{2}\tilde{\mu}_1^{pp}\phi_a}{1 + \frac{1}{2}\tilde{\mu}_1^{pp}\phi_a} \quad \text{(S4)}$$

Where $\tilde{\mu}_1^{pp} = \frac{12\eta\mu^{pp}}{\pi H^3 d^2}$ (Eq.45, reference 5) and $\mu^{pp}$ is the polarizability coefficient for a droplet. However, the $\mu^{pp}$ is not known, but we can estimate an average value from $\bar{v}^{pp}$ using Eq. 86b from reference 5

$$\bar{v}^{pp} = H\ \bar{v}_0^{pp}(1 - 12\ \bar{n}_s H^{-3}\eta\ \bar{\mu}^{pp}) \quad \text{(S5)}$$

Where $\bar{\mu}^{pp}$ is the average polarizability coefficient and $\bar{n}_s$ is the number density of the droplets.

Substituting the Eq. of S5 in equation S4, results in:

$$\tilde{\mu}_1^{pp} = \frac{H\bar{v}_0^{pp} - \bar{v}^{pp}}{\bar{n}_s H \bar{v}_0^{pp}} \cdot \frac{1}{\pi d^2} \quad \text{(S6)}$$

By substituting the value of $\bar{v}^{pp}$ and $\bar{n}_s$ from experiment with lowest area fraction, we find $\tilde{\mu}_1^{pp} = 3.73$.

## 5 - Porous medium approximation to explain the effect of area fraction on $\chi_c$

From the plots in Fig. 4a and b, it can be observed that the convective transport coefficient $\chi_c$ shows an increasing trend with respect to the particle density $\bar{\phi}_a$. To provide a qualitative explanation we consider a simple model where the particle phase is treated as a porous medium that slides through the channel due to the pressure gradient $\nabla P$ and is restricted by friction from the sidewalls. The fluid phase penetrates through the particle phase with a permeability $\mu$.

The relative velocity of the fluid phase with respect to the particles can be expressed by Darcy's law as

$$V - U = -\mu \nabla P \qquad \text{(S7)}$$

where *V* and *U* are the fluid and particle phase velocities. While the balance between the pressure force $F_p = -hwL\nabla P$ and the friction force $F_{wf} = wLfU$ responsible for the motion of the porous particle phase gives an expression for the pressure gradient

$$\nabla P = -\frac{fU}{h} \qquad \text{(S8)}$$

Combining eqn. S7 and S8 we obtain the following relation for relative fluid velocity:

$$V - U = \bar{f}U \qquad \text{(S9)}$$

where $\bar{f} = f\mu/h$ is the dimensionless friction factor.

Assuming a cylindrical shape of droplets due to strong confinement, the velocity of the suspension can be expressed as a weighted sum of the particle phase and fluid phase velocities as

$$u = \bar{\phi}_a U + (1 - \bar{\phi}_a)V = U + (1 - \bar{\phi}_a)\Delta V \qquad \text{(S10)}$$

Using eqn. S9, the above equation can be expressed as a relation between the particle phase velocity u, suspension velocity U, the dimensionless friction factor $\bar{f}$:

$$u = U\left[1 + (1 - \bar{\phi}_a)\bar{f}\right] \qquad \text{(S11)}$$

Finally, by using eqn. S11 and the definition for convective transport coefficient Eq. 3-3, we arrive at an expression which directly relates the convective transport coefficient $\chi_c$ to the area-fraction of particles $\bar{\phi}_a$ in the suspension and the dimensionless friction factor

$$\frac{1}{\chi_c} = \left[1 + (1 - \bar{\phi}_a)\bar{f}\right] \qquad \text{(S12)}$$

From eqn S12 one can observe that as $\bar{\phi}_a \to 1$, $\chi_c \to 1$ which emphasizes the impenetrability of the particle phase to fluid flow at very high densities. For intermediate values, one can realize that when area fraction $\bar{\phi}_a$ increases, the total number of particles in contact with the sidewall also increases leading to an increase in the friction factor $f$. However, due to the deformability of drops the permeability of the medium falls drastically and at a rate higher than the increase of the friction factor. Additionally, the factor $(1 - \bar{\phi}_a)$ also decreases with density. Hence, eqn. S12 decreases strongly with particle density.

## 6 – Stabilization of the dense-droplet band in stronger confinement

We have conducted experiments in the confinement ratio of $d/h = 0.8$ for spherical droplets. We have observed that a dense closely packed droplet band can prevent formation of fingers at this stronger confinement. This is due to the non-zero yield-stress of the ordered emulsion:

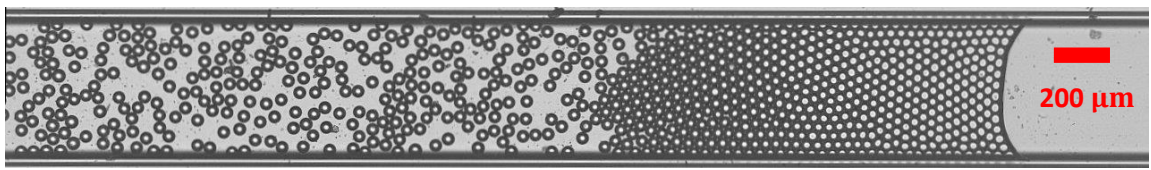

**Fig. S7 Presence of a closely-packed droplet-band under strong confinements can prevent fingering instability**

## Supplementary references